\documentstyle[epsf,aps]{revtex}
\newcommand{\mafigura}[4]{
  \begin{figure}[hbtp]
    \begin{center}
      \epsfxsize=#1 \leavevmode \epsffile{#2}
    \end{center}
    \caption{#3}
    \label{#4}
  \end{figure} }
\newcommand{\eq}[1]{eq.~(\ref{#1})}
\newcommand{\as}{\alpha_s}
\newcommand{\api}{\frac{\alpha_s}{\pi}}
\newcommand{\yc}{y_{c}}
\newcommand{\rb}{r_b}
\newcommand{\el}[1]{^{(#1)}}
\begin{document}
\preprint{FTUV/96-80}
\preprint{IFIC/96-89}
\preprint{PRA-HEP 97/4}

\title{Do the quark masses run? Extracting $\bar{m}_b(m_Z)$ from LEP data}

\draft

\author{Germ\'an Rodrigo and Arcadi Santamaria}
\address{Departament de F\'{\i}sica Te\`orica, IFIC,
CSIC-Universitat de Val\`encia, 46100 Burjassot, Val\`encia, Spain} 
\author{Mikhail Bilenky\thanks{On leave from JINR, Dubna , Russian Federation}}
\address{Institute of Physics, AS CR, 18040 Prague 8, and 
Nuclear Physics Institute, AS CR, 25068 \v{R}e\v{z}(Prague), Czech Republic}

\date{March 18, 1997}

\maketitle

\begin{abstract}
We present the first results of next-to-leading order QCD corrections
to three jet heavy quark production at LEP including mass effects. Among
other applications, this calculation can be used to extract the bottom quark
mass from LEP data, and therefore to test the running of masses as
predicted by QCD.
\end{abstract}


The decay width of the $Z$ gauge boson into three jets has already been 
computed at the leading order (LO) including complete 
quark mass effects 
\cite{ballestrero.maina.ea:92,ballestrero.maina.ea:94,bilenky.rodrigo.ea:95}
where it has been shown that 
mass effects could be as large as 1\% to 6\%,
depending on the value of the mass and the jet resolution parameter
$\yc$. In fact, these effects
had already been seen in the experimental tests of the flavor
independence of the strong coupling constant 
\cite{chrin:93,javalls:94,fuster:94,p-abreu-et-al:93,d-buskulic-et-al:95}. 
In view of that we 
proposed \cite{bilenky.rodrigo.ea:95}, together with the DELPHI collaboration 
\cite{fuster:93},
the possibility of using the ratio 
\cite{bilenky.rodrigo.ea:95,fuster:94,p-abreu-et-al:93}
\begin{equation}
R^{bd}_3 \equiv \frac{\Gamma^b_{3j}(\yc)/\Gamma^b}
{\Gamma^d_{3j}(\yc)/\Gamma^d}
\label{eq:r3bd_def}
\end{equation}
as a means to extract the bottom quark mass from LEP data.
In this equation $\Gamma^q_{3j}(\yc)/\Gamma^q$ is the three-jet fraction
of $Z$ decays into the quark $q$ and
$\yc$ is the jet resolution parameter. 

Since the measurement of $R^{bd}_3$ is done
far away from the threshold of $b$ quark production, it will allow, 
for the first time, to test the running of a quark mass as predicted by QCD.
However, in \cite{bilenky.rodrigo.ea:95} we also discussed that
the leading order calculation does not distinguish among the
different definitions of the quark mass, perturbative pole mass, $M_b$,
running mass at $M_b$, or running mass at $m_Z$. Therefore in order
to correctly take into account  mass effects it is necessary to 
perform a complete next--to--leading order (NLO) calculation of three
jet ratios including quark masses \cite{rodrigo:96,rodrigo:96*b,santamaria:96}.

In this letter we sketch the main points of this calculation,
leaving the details of the complete calculation for other
publications \cite{rodrigo.santamaria.ea:97*b,rodrigo.santamaria.ea:97*c}, 
and we present the results that have been used by the DELPHI collaboration 
to measure the running mass of the bottom quark at $\mu=m_Z$ 
\cite{fuster.cabrera.ea:96,fuster:97}.

In the last years the most popular definitions of jets
are based on the so-called jet clustering algorithms. 
These algorithms can be applied at the parton
level in the theoretical calculations and also to the
bunch of real particles observed at experiment.
In the jet-clustering algorithms jets are defined as follows:
starting from  a bunch of particles
with momenta $p_i$ one computes, for example, a quantity like
$y_{ij}=2 \min(E_i^2, E_j^2)/s\ (1-\cos \theta_{ij})$
for all pairs $(i,~j)$ of particles. Then one takes the minimum of
all $y_{ij}$ and if it satisfies that it is smaller than a given quantity
$\yc$ (the resolution parameter, y-cut)  the two particles
which define this $y_{ij}$ are regarded
as belonging to the same jet, therefore, they are recombined into a new
pseudoparticle by defining the four-momentum of the pseudoparticle according
to some rule, for example, $p_k = p_i + p_j$.
After this first step one  has a bunch of pseudoparticles and
the algorithm can be applied again and again until all the pseudoparticles
satisfy  $y_{ij} > \yc$. 
The number of pseudoparticles found in the end
is the number of jets in the event.
This procedure leads automatically to IR finite quantities 
because one excludes the regions of phase space that cause trouble.
It has been shown that, for some of the algorithms,
the passage from partons to hadrons (hadronization)
does not change much the behavior of the observables 
\cite{bethke.kunszt.ea:92},
thus allowing to compare theoretical predictions
with experimental results. 

Although we have studied the four jet-clustering algorithms
discussed in 
\cite{bilenky.rodrigo.ea:95,rodrigo:96*b,bethke.kunszt.ea:92,kunszt.nason.ea:89},
here we will present results only for the DURHAM algorithm 
\cite{brown.stirling:92*b,catani.dokshitser.ea:91} 
which is the one we just have defined
and seems to be the one that behaves better for most of the observables.

The decay width of the $Z$ boson into three jets containing the bottom
quark mass can be written as follows \cite{bilenky.rodrigo.ea:95}
\begin{equation}
\Gamma^b_{3j}(\yc) = m_Z \frac{g^2}{c_W^2 64 \pi} \frac{\alpha_s(m_Z)}{\pi}
\left( g_V^2 H_V(\yc,\rb)+g_A^2 H_A(\yc,\rb)\right)~,
\label{eq:gamma3jets}
\end{equation}
where $g$ is the SU(2) gauge coupling constant,
$c_W = \cos \theta_W$ and $s_W = \sin\theta_W$
are the cosine and the sine of the weak mixing angle,
$g_V =-1+4/3 s_W^2$ and $g_A=1$ are the vector and axial coupling of the $Z$ 
boson to the bottom quark, and $H_V(\yc,\rb)$ and $H_A(\yc,\rb)$ 
contain
all the dependences in the jet resolution parameter, $\yc$, and the quark
mass, $\rb = (M_b/m_Z)^2$, for the vector and axial parts in the different
algorithms. These functions can be expanded in $\as$, and, if the leading 
dependence on the quark mass is factorized out we have,
\begin{eqnarray}
H_{V(A)}(\yc,\rb) &=& A\el{0}(\yc)+\api A\el{1}(\yc)+ \nonumber \\
&& \rb\left( B_{V(A)}\el{0}(\yc,\rb)
+\api B_{V(A)}\el{1}(\yc,\rb)\right) + \cdots~.
\label{eq:hexpansion}
\end{eqnarray}
Here,  $A\el{0}(\yc)$ is the tree level contribution in the case of
massless quarks and it is known for the different algorithms
in analytic form. It is exactly the same function for the vector and the axial
parts. The functions $B_{V(A)}\el{0}(\yc,\rb)$ take into account tree level 
mass effects
once the leading dependence in $\rb$ has been factorized out. 
They were calculated numerically in \cite{bilenky.rodrigo.ea:95} for the 
different algorithms and results were presented in the form of fits to
the numerical results.  
$A\el{1}(\yc)$ gives the QCD next--to--leading order correction in the
case of massless quarks and to good approximation it is the same for the
vector and the axial parts [Because of the triangle  anomaly there are 
one-loop triangle diagrams 
contributing to $Z\rightarrow \bar{q}\bar{q} g$  with the top and the 
bottom quarks running in the loop. Since $m_t \not= m_b$ the anomaly 
cancellation is not complete. These diagrams contribute to the axial part 
even for $m_q=0$ and  lead to a deviation from 
$A\el{1}_V(\yc) = A\el{1}_A(\yc)$
\cite{hagiwara.kuruma.ea:91}. 
This deviation is, however, small \cite{hagiwara.kuruma.ea:91} and we are
not going to consider its effect here]. This function is also known
for the different algorithms
\cite{bethke.kunszt.ea:92,kunszt.nason.ea:89}. 
Finally, the functions $B_{V(A)}\el{1}(\yc,\rb)$ were completely unknown
and contain the next-to-leading order corrections depending on the
quark mass.  In the next section we present our results in the form
of fits to some combinations of the relevant functions in the case of
the DURHAM algorithm, which is the one that gives smaller radiative
corrections, and we postpone the presentation of results for the different
functions and  algorithms for another publication 
\cite{rodrigo.santamaria.ea:97*c}.
Note that the way we write $H_{V(A)}(\yc,\rb)$ in
\eq{eq:hexpansion} is not an expansion for small $\rb$. We keep
the exact dependence on $\rb$ in the functions  $B_{V(A)}\el{0}(\yc,\rb)$
and $B_{V(A)}\el{1}(\yc,\rb)$. Factoring out $\rb$  makes it
easier to analyze the massless limit and the dependence of the results
on $\rb$ in the region of interest. This means that
our results can also be adapted, by including the photon exchange, to compute
the $e^+ e^-$ cross section into three jets out of the $Z$ peak at 
lower energies or at higher energies in top quark production.

At the NLO we have contributions to the three-jet cross section from three
and four parton final states. One loop three-parton amplitudes are both IR
and UV divergent.  Therefore, some regularization procedure is needed. We use
dimensional regularization 
for both IR and UV divergences because
dimensional regularization preserves the QCD Ward identities. The UV 
divergences,
however, can be easily removed by renormalization since the appropriate
counterterms are very well known. 
The three-jet cross section is obtained by integrating both contributions,
renormalized three-parton and four-parton amplitudes, in
the three-jet phase space region defined by the different 
jet clustering algorithms. This quantity is
infrared finite and well defined, although the three and the four parton
transition amplitudes independently contain infrared singularities.

The three-parton transition amplitudes can be expressed in terms of a few
scalar one-loop integrals. The result contains poles in $\epsilon=(4-D)/2$,
where $D$ is the number of space-time dimensions. Some of the poles
come from UV divergences and the other come from IR divergences.
After UV renormalization we obtain
analytical expressions for the terms proportional to the infrared poles and
for the finite contributions. The infrared poles will cancel against
the four-parton contributions. 
The finite contributions are integrated
numerically in the three-jet region. 

The four-parton transition amplitudes are
split into a soft and collinear part in the three-jet region and a hard 
contribution. The
soft terms are integrated analytically in arbitrary $D$
dimensions in the region of the phase space containing the infrared
singularities. We obtain analytical expressions for the infrared behavior
of the four-parton transition amplitudes and we show how these infrared
terms cancel exactly the infrared singularities of the three-parton
contributions. The hard terms are calculated in $D=4$ dimensions.
The remaining phase space integrations, giving rise to finite contributions,
are performed numerically. 

Following Ellis, Ross and Terrano \cite{ellis.ross.ea:81} we have classified
both, three-parton and four-parton transition probabilities, according to 
their color factors. It is clear that the cancellation of IR divergences 
between
three-parton and four-parton processes can only occur inside groups of
diagrams with the same color factor. The cancellation of IR
divergences can be seen more clearly by representing the different
amplitudes as the different cuts one can perform in the three-loop bubble
diagrams contributing to the $Z$-boson selfenergy. 
After summing up the three-parton and four-parton contributions to the
three-jet decay width of the $Z$ boson we obtain the functions
$H_V(\yc,\rb)$ and $H_A(\yc,\rb)$ in \eq{eq:gamma3jets} at order
$\as$. Details of the calculation, cancellation of
divergences and results for the relevant functions will be presented
elsewhere \cite{rodrigo.santamaria.ea:97*b,rodrigo.santamaria.ea:97*c}.
Since a large part of the calculation has been done numerically, 
it is important to have some checks of it. We have performed the following 
tests:

\begin{itemize}
\item We have checked our four parton amplitudes in the massless limit
against the amplitudes presented by Ellis-Ross-Terrano (ERT)
\cite{ellis.ross.ea:81}.
The three-parton amplitudes for massive quarks cannot be compared directly 
with the massless results because 
collinear poles in $\epsilon$, in the massless case, appear 
as logarithms of the quark mass when the mass is taken into account.
\item The four parton transition amplitudes have also been checked in the 
case of massive quarks, in four dimensions, by comparing their contribution 
to four jet processes to the known results \cite{ballestrero.maina.ea:92}
\item Of course we have checked that, also in the massive case,
all the IR divergences cancel between
three parton and four parton contributions
\cite{bloch.nordsieck:37,kinoshita:62,lee.nauenberg:64}.
\item To check the performance of the numerical programs and the overall
approach we have applied our method to the massless amplitudes of ERT and 
obtained the known results for the functions $A\el{1}$. 
\item We have checked, independently for each of the groups of diagrams
with different color factors, that the final result obtained with massive 
quarks reduces to the massless result in the limit of very small masses.
\end{itemize}

The last test is the main check of our calculation.
We have calculated the functions
$H_V(\yc,\rb)$ and $H_A(\yc,\rb)$ for several small values of
$\rb$, in the range $M_b\sim 1 - 5~GeV$, and then we have extrapolated 
the results for $\rb \rightarrow 0$. 
In that limit we recover the values for the function
$A\el{1}(\yc)$ in the different algorithms considered and the different
groups of diagrams.
This check is not trivial at all
since the structure of IR divergences for massive quarks
is quite different from the case of massless quarks: for massive quarks
collinear divergences are regulated by the quark mass, and therefore some of 
the poles in
$\epsilon$ that appear in the massless case are softened by
$\log \rb$. 
Although these checks do not constitute a complete test of the massive 
corrections, we think that all together give some kind of confidence in the 
final result for massive quarks.

Now, to obtain $R^{bd}_3$, we can substitute \eq{eq:hexpansion} in 
\eq{eq:gamma3jets}, this into \eq{eq:r3bd_def} and use the
value of $\Gamma^b$ which is also well known (for
mass effects at order $\as$ see for instance 
\cite{bilenky.rodrigo.ea:95,djouadi.kuhn.ea:90,chetyrkin.kuhn:90}).
Putting everything together and expanding in $\as$ and $\rb$ we can 
write
\begin{equation}
R^{bd}_3 = 1+ \rb \left(b_0(\yc,\rb)+\frac{\as}{\pi} b_1(\yc,\rb)\right)~,
\label{eq:r3resultpole}
\end{equation}
where the functions $b_0$ and $b_1$ are an average of the
vector and axial parts, weighted by
$c_V= g_V^2/(g_V^2+g_A^2)$ and
$c_A= g_A^2/(g_V^2+g_A^2)$ respectively, and can be written in terms of
the different functions introduced before, \eq{eq:hexpansion},  
\cite{bilenky.rodrigo.ea:95,rodrigo:96*b}

It is important to note that because the particular normalization we have
used in the definition of $R^{bd}_3$, which is manifested in the final 
dependence
on $c_V$ and $c_A$, most of the electroweak corrections cancel. Those are
about 1\% \cite{bernabeu.pich.ea:91}
in total rates while in $R^{bd}_3$ are below 0.05\%. Therefore
for our estimates it is enough to consider tree level values of $g_V$ and
$g_A$. The same argument applies for the passage from decay widths to
cross sections. Contributions from photon exchange are small at LEP and
can be absorbed in a redefinition of $g_V^2$ and $g_A^2$ 
\cite{jersak.laermann.ea:82}.
They will add a small correction to our observable.

Although intermediate calculations have been performed using the 
pole mass we can also re-express our results in terms of the running
quark mass by using the known perturbative expression
\begin{equation}
M_b^2 = \bar{m}_b^2(\mu) \left[1+2\frac{\as(\mu)}{\pi}
\left(\frac{4}{3}-\log \frac{m_b^2}{\mu^2} \right)\right]~.
\label{eq:poltorunning}
\end{equation}
The connection between pole and running masses is known up to
order $\as^2$, however consistency of our pure perturbative 
calculation requires we use only the expression above. We obtain
\begin{equation}
R^{bd}_3 = 1+ \bar{r}_b(\mu) \left(
b_0(\yc,\rb)+\frac{\as(\mu)}{\pi} 
\left(\bar{b}_1(\yc,\rb)-2 b_0(\yc,\rb) \log \frac{m_Z^2}{\mu^2}\right)
\right)~.
\label{eq:r3resultrunning}
\end{equation}
Where $\bar{r}_b(\mu)=\bar{m}^2_b(\mu)/m_Z^2$ and
\begin{equation}
\bar{b}_1(\yc,\rb) = b_1(\yc,\rb)+b_0(\yc,\rb)\left[8/3-2\log(\rb)\right]~.
\label{eq:b1bar}
\end{equation}
$\bar{r}_b(\mu)$ can be expressed in terms of the running mass of
the $b$ quark at $\mu = m_Z$ by using the renormalization group. At the
order we are working 
\begin{equation}
\bar{r}_b(\mu)=\bar{r}_b(m_Z)\left(
\frac{\alpha_s(m_Z)}{\alpha_s(\mu)}\right)^{-4\gamma_0/\beta_0}
\ \ \ \ {\mathrm{with}}\ \ \ \
\alpha_s(\mu) = \frac{\alpha_s(m_Z)}{1+\alpha_s(m_Z)\beta_0 t} 
\label{eq:mbrunning}
\end{equation}
and $t=\log(\mu^2/m^2_Z)/(4\pi)$, $\beta_0 =11-2N_f/3$, $N_f=5$ 
and $\gamma_0=2$.

At the perturbative level \eq{eq:r3resultpole} and \eq{eq:r3resultrunning}
are equivalent. However, since they neglect different higher order terms
they lead to different answers. Since the experiment is performed at
high energies (the relevant scales are $m_Z$ and $m_Z \sqrt{\yc}$) one
would think that the expression in terms of the running mass is more
appropriate because the running mass is a true short distance parameter
while the pole mass contains in it all the complicated physics at
scales $\mu \sim M_b$. Moreover, by using the expression in terms of the
running mass we can vary the scale in order to estimate the error
due to the neglect of higher order corrections.
In any case, if one  would push the
result of \eq{eq:r3resultrunning} up to scales as low as $\mu = 5~GeV$
one would get something closer to the pole mass result. Therefore,
we use \eq{eq:r3resultrunning} and vary the scale 
in an appropriate range to obtain an estimate of the error in the calculation.

The function $b_0(\yc,\rb)$ gives the mass corrections at leading order. 
As shown
in \cite{bilenky.rodrigo.ea:95} it depends very mildly on 
the quark mass
in the region of interest ( $M_b \sim 3 - 5~GeV$). Therefore it is appropriate
to present our results for $b_0(\yc,\rb)$  as a fit 
in only $\yc$: $b_0(\yc,\rb) = \sum_{n=0}^2 k_0^{(n)}$ log${}^n y_c$.
For the DURHAM algorithm,
in the range $0.01 < y_c < 0.10$ and $3~GeV < M_b < 5~GeV$ and using
$\sin^2 \theta_W = 0.2315$ we obtain
$k_0^{(0)}=-10.521\;$,  $k_0^{(1)}=-4.4352 \;$, $k_0^{(2)}=-1.6629\;$.

The function $\bar{b}_1(\yc,\rb)$ is the main result of this paper.
It gives the NLO  massive
corrections to our observable. It is important to note that $\bar{b}_1$
contains significant logarithmic corrections depending on the quark mass. They
come from different sources: first, the NLO corrections written in terms of the
pole mass now contain some residual mass dependence, second the normalization
to the total rate induces some additional logarithmic dependences,
and finally the passage from the pole mass to the running mass
adds also an additional logarithmic dependence. Therefore,
we choose to include explicitly this dependence on the quark mass in our
fits to the function $\bar{b}_1(\yc,\rb)$. For our purposes a fit of the
form
$\bar{b}_1(\yc,\rb)= k_1^{(0)} + k_1^{(1)}\log(\yc) +
k_m^{(0)} \log (\rb)$
is good enough.
The coefficients we obtain for the DURHAM algorithm and in the ranges
we just mentioned for $\yc$ and $\rb$ are:
$k_1^{(0)}=297.92\;$,  $k_1^{(1)}=59.358 \;$, $k_m^{(0)}=46.238\;$.

In fig.~\ref{fig:r3} we present $R_3^{bd}$
for $\mu=m_Z$ (dashed), $\mu=30~GeV$ (dashed-dotted) and $\mu= 10~GeV$
(dotted) for $\bar{m}_b(m_Z)= 3~GeV$ and $\alpha_s(m_Z)=0.118$.
For comparison we also present the LO results for
$M_b = 5~GeV$ (lower solid line) which is, roughly,
the value of the pole mass obtained at low energies and
$\bar{m}_b(m_Z) = 3~GeV$ (upper solid line)
which is, roughly, the value one obtains for the running mass at the $m_Z$
scale by using the renormalization group \cite{rodrigo:95}. 
Note that choosing
a low value for $\mu$ makes the result closer to the LO result written
in terms of the pole mass, while choosing a large $\mu$ makes the result
approach to the LO result written in terms of the running mass at the
$m_Z$ scale. 
If $R_3^{bd}$ is measured to good accuracy one could use
\eq{eq:r3resultrunning}
and \eq{eq:mbrunning} to extract $\bar{m}_b(m_Z)$. However, 
the extracted
result will depend on $\mu$.
For illustration, in fig.~\ref{fig:mbbar} we represent, as a
function of $\mu$, the value one would obtain for $\bar{m}_b(m_Z)$ if
$R^{bd}_{3\, exp}=0.96$ for $\yc = 0.02$. 
What is the best scale one should
choose in three-jet quantities is a long standing discussion. One would
think that if the energy is equally distributed among the three jets one
should choose $\mu \sim m_Z/3$. A conservative approach is to vary the
scale in an appropriate range and take the spread of the result as
an estimate of the error due to higher order corrections.
From fig.~\ref{fig:mbbar} we see that if we take $\mu$ in the range
$m_Z/10 - m_Z$ the error due to the scale and $\as$ in the determination
of $\bar{m}_b(m_Z)$ would be of about $0.20~GeV$. If scales as low as
$\mu = 5~GeV$ are accepted the error increases to $0.23~GeV$.
Whether this error can be reduced by a clever choice of the scale or
resummation of leading logs in $\yc$ or $\rb$ remains to be seen.

The calculation presented in this paper has already been used by
the DELPHI collaboration \cite{fuster:97} to extract $\bar{m}_b(m_Z)$. 
The preliminary result [we like to thank the DELPHI collaboration for
allowing us to use these numbers here] they obtain is
\begin{equation}
\bar{m}_b(m_Z) =  2.85 \pm 
0.22~({\mathrm stat})\pm 0.20~({\mathrm theo}) 
\pm 0.36~({\mathrm fragmentation})~GeV  
\label{eq:mbmeasure}
\end{equation}
which has to be compared with low energy determinations of the bottom quark
mass. The last analysis of the $\Upsilon$ system using QCD sum rules 
\cite{jamin.pich:97} gives
$\bar{m}_b(\bar{m}_b) = 4.13 \pm 0.06~GeV$
which translates into $\bar{m}_b(m_Z) = 2.83 \pm 0.10~GeV$ if one uses
three loop renormalization group running and $\alpha_s(m_Z)=0.118\pm 0.003$. 
On the other hand,
the last lattice result gives\cite{gimenez.martinelli.ea:97}
$\bar{m}_b(\bar{m}_b)=4.15 \pm 0.20~GeV$ and 
$\bar{m}_b(m_Z) = 2.84 \pm 0.21~GeV$.
Given the errors it is clear that central values agree so well just by chance. 
In particular the value
in \eq{eq:mbmeasure} has been extracted by using a different central value
for $\alpha_s$.

It is encouraging to see that this preliminary measurement is in full
compatibility with low energy data, and, although for the moment it
is not competitive with low energy measurements,
it is good enough for testing the
running of the bottom quark mass from $\mu=M_b$ to $\mu=m_Z$: the
result for $\bar{m}_b(m_Z)$ in \eq{eq:mbmeasure} and the
previous values for $\bar{m}_b(\bar{m}_b)$ differ 
by more than 2.5 standard deviations. 
We believe that these results can be substantially improved with more
experimental and theoretical work.

\vspace{1cm}

We would like to acknowledge interesting discussions with 
S. Catani, V. Gim\'enez, M. Jamin, H. K\"uhn, A. Manohar, G. Martinelli, 
S. Narison and A. Pich. We are also indebted with S. Cabrera, 
J. Fuster and S. Mart\'{\i} for an
enjoyable collaboration. 
M.B. thanks the Univ. de Val\`encia for the warm hospitality during
his visit.
The work of G.R. and A.S. has been supported in
part by CICYT (Spain) under the grant AEN-96-1718 and IVEI.
The work of G.R. has also been supported in part by CSIC-Fundaci\'o Bancaixa.

\vspace{1cm}

The results in this paper have been previously presented at several
conferences\cite{rodrigo:96,fuster.cabrera.ea:96,santamaria:96,fuster:97}
and were submitted as part of the Ph.D. requirements for G.R.. In the
meanwhile a preprint dealing with the same problem has 
appeared\cite{bernreuther.brandenburg.ea:97}.

\vfil\eject
\mafigura{8 cm}{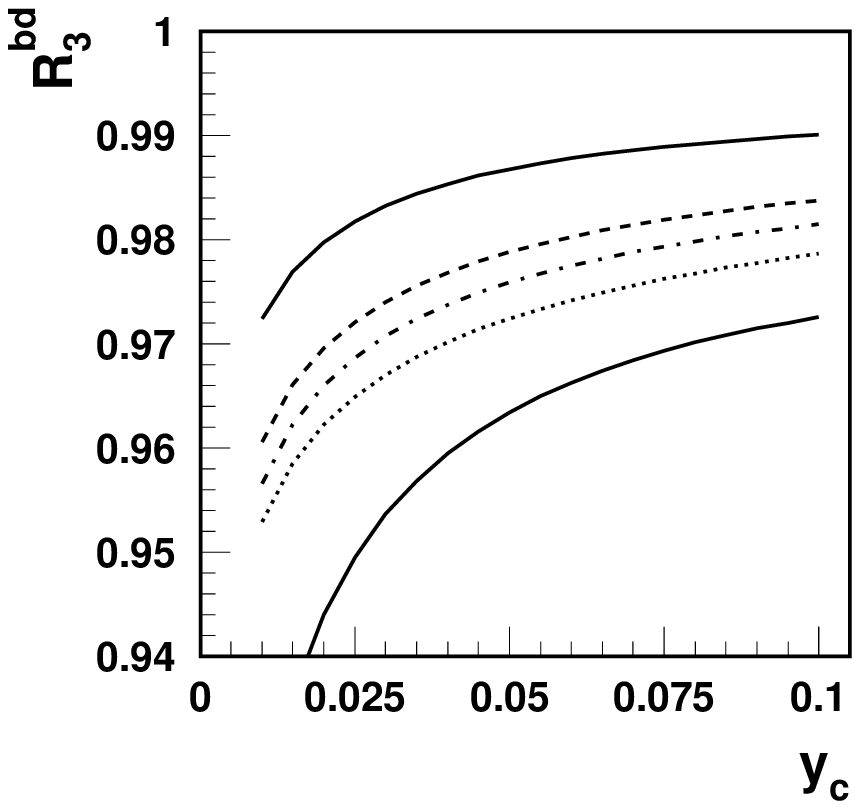}{
NLO results for $R_3^{bd}$ for $\mu=m_Z$ (dashed), 
$\mu=30~GeV$ (dashed-dotted) 
and $\mu= 10~GeV$ (dotted) for $\bar{m}_b(m_Z)= 3~GeV$ and 
$\alpha_s(m_Z)=0.118$. For comparison we also plot the LO results for
$M_b = 5~GeV$ (lower solid line) and $\bar{m}_b(m_Z) = 3~GeV$ 
(upper solid line)
}{fig:r3}
\mafigura{8 cm}{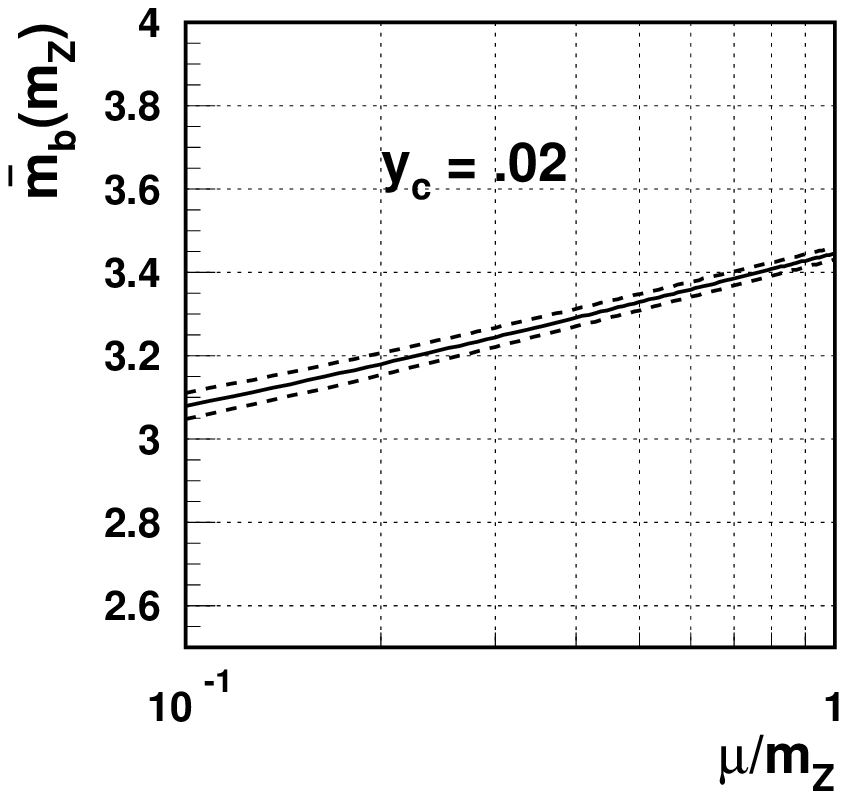}{
Extracted value of $\bar{m}_b(m_Z)$ if $R^{bd}_{3\, exp}=0.96$ as
a function of the scale $\mu$. We take $\alpha_s(m_Z)=0.118$ (solid)
and $\Delta \as = 0.003$ (dashed).
}{fig:mbbar}
\end{document}